\documentclass[pre,aps,preprint,showpacs]{revtex4} 
\usepackage{revsymb,amsmath}
\usepackage{graphicx}
\begin{document}
\author{Beatrix M. Schulz, Steffen Trimper}
\affiliation{Fachbereich Physik,
Martin-Luther-Universit\"at,D-06099 Halle, Germany}
\email{trimper@physik.uni-halle.de}
\author{Michael Schulz}
\affiliation{Abteilung Theoretische Physik, Universit\"at Ulm, D-89069 Ulm, Germany}
\email{michael.schulz.@physik.uni-ulm.de}
\title{Spin facilitated Ising model with long range interaction}
\date{\today }

\begin{abstract}
We study the dynamics of a spin facilitated Ising model with long range kinetic constraints. To formulate 
those restrictions within an analytical approach we introduce the size of a kinetic active environment of a 
given spin. Based on a Master equation in second quantized form, the spin-autocorrelation function is calculated. It exhibits a pronounced slow dynamics, manifested by a logarithmic decay law of the spin-autocorrelation function. 
In case of an infinite kinetic interaction the mean field solution yields an asymptotic exact expression for the autocorrelation function which is in excellent agreement with Monte Carlo Simulations for finite interaction lengths. 
With increasing size of the active zone the cooperative processes, characterizing the facilitated model with short 
range kinetic interaction, become irrelevant. We demonstrate that the long range kinetic interaction dominates the 
actual spin configurations of the whole system and the mean field solution is the exact one.  
\end{abstract}
\maketitle

\section{Introduction}

A central topic in non equilibrium statistical physics is the quantitative and qualitative understanding of long-time 
phenomena in strongly interacting many-body systems without long range order. Such systems are characterized by a 
whole hierarchy of degrees of freedom \cite{psaa}, which generate for example a broad distribution of relaxation times. 
The different levels relax in series resulting in an extreme slowing-down of dynamical processes. These processes and 
other singular behavior is in contrast to the critical slowing-down observed in the vicinity of conventional phase 
transitions. As a consequence of the mentioned hierarchy strongly disordered systems display a high degree of 
cooperativity of local processes. Thus, a signature for a broad class of dense systems are autocorrelation functions that 
can be fitted by a stretched exponential decay law for sufficiently long times or by a non-Arrhenius temperature 
dependence of relaxation times. Although it remains an open problem whether these phenomena are universal,  
they are, at least, a characteristic feature for many systems, in particular, when the systems are far from thermal 
equilibrium. For instance, such scenarios arise for the main glass transition process in supercooled liquids.\\ 
\noindent One way to model the hierarchy of relaxation processes is given by kinetic constraints. Due to that constraint 
a local dynamical process is free to evolve after changing its environment. A rather simple realization is given by 
facilitated kinetic spin systems which offer the typical slowing-down of dynamical processes, caused by an 
increasing cooperativity of local spin-flip processes with decreasing temperature. In the present paper our interest is focused on the analysis the f-spin facilitated  kinetic Ising models \cite{Fredrickson1,Fredrickson2,Fredrickson3}, originally introduced by Fredrickson and 
Andersen. These models are formulated on $d$-dimensional lattices. To each lattice site $i$ we assign a spin variable $\sigma _i$ with two possible states $\sigma _i = \pm 1$. The total set of all observables 
${\bf \sigma } = \left\{\sigma _i \right\}$ forms a configuration. The basic dynamic of the facilitated kinetic Ising model is given by stochastic 
spin-flip processes $\sigma_i = +1 \leftrightarrow \sigma _i = -1 $. Additional to these stochastic Glauber 
dynamics \cite{Glauber} the system is subjected to self-induced topological restrictions. The origin of the imposed constraints  becomes evident when one considers the local spin variable $\sigma _i$ as coarse-grained local density variable. 
The down-state is associated to a low-density region while the up-state represents a 
high-density region. Whereas the local stochastic spin - flip processes are controlled by the thermodynamic Gibbs measure, the constraints reflect the mentioned cooperativity arising at low temperatures. An elementary spin-flip process 
at lattice point $i$ is only possible after an appropriate rearrangement of the spins in the neighborhood of the 
site $i$. In other words, a given spin $\sigma _i$ has to wait until the degrees of freedom in its environment have reached a certain configuration specified below. Insofar, the kinetic restrictions are related to the hierarchy of degrees of freedom \cite{se}.\\ 
\noindent To be specific, the restrictions are explicitly taken into account by choosing the local flip rates depending 
on the orientation of its neighbors. A spin at the lattice point $i$ can only undergo a flip process in case the number 
$z$ of nearest neighbored lattice point in an up state $\sigma _j = +1\quad j=1 \dots z$ is equal or exceed the 
restriction number $f$ with $0 < f < z$. Such kind of models \cite{Fredrickson1,Fredrickson2,Fredrickson3,Fredrickson4} are denoted as f-spin facilitated Ising models. 
For a realization on a d-dimensional lattice we use the abbreviation SFM[f,~d]. Notice that in the original version 
the SFM[f,~d] comprises a system of noninteracting Ising spins. A coupling is realized only via the kinetic constraints. 
Identifying the spin variables as local density variables as done above, the restrictions lead inevitably to 
cooperative rearrangements of the whole spin systems. Due to the kinetic constraints glassy effects arise at low temperatures \cite{schu1,schu2,schu3,schu4}.\\  
\noindent The SFM[f,d] allows also another interpretation in terms of chemical reactions. To that aim let us identify 
the spin-up state as a molecule of type $A$ and the down-state as a molecule of type $B$, respectively. Then the spin facilitated model can be considered as a reversible auto catalytic chemical reaction $f A + B \rightleftharpoons (f + 1) A$ on a d-dimensional lattice. The molecule changes its constitution from $A$ to $B$ and/or from $B$ to $A$ only in case, 
when at least $f$ molecules of type $A$ are situated in the nearest environment.\\      
\noindent Obviously the dynamics of the SFM[f,~d] may not only determine by the restriction number $f$, but also by the 
range of the kinetic active environment of a certain spin $\sigma _i$. In particular, the long time behavior of the SFM should be changed significantly, when the number of spins in the neighborhood is enlarged. One aim of the paper 
is to extend the original version of the spin facilitated Ising model in such a manner that a long range interaction 
will be included into the model. To realized a minimal extension of the SFM we define a quantity $E_i(R)$ as the 
environment of an arbitrary spin $\sigma _i$ situated at the lattice point $i$. Here, the new parameter $R$ is, so to say, the radius of the neighborhood. The simplest assumption consists of that 
$E_i(R)$ contains all the spins $\sigma _k$, the distance of which to the spin $\sigma _i$ fulfill the condition 
$\mid k - i \mid \leq R$. The total number of spins within that region of the radius $R$ is denoted by $G$ where 
$G \geq f$. Thus, if at least a subset of $f$ spins in the region $E_i(R)$ is in the up state $+1$, the spin $\sigma _i$ 
is able to flip. As a consequence we expect a more pronounced slowing down of the relaxation processes with an 
increasing ratio $f/G$ with fixed $f$ as well as for increasing $f$ with fixed ratio $f/G$. Indeed, we demonstrate 
an utmost drastic slowing down, namely a logarithmic decay of the autocorrelation function.

\section{Analytical investigations}

\subsection{Evolution equations}

Let us start with a brief review of the main ideas behind the Fock--space formalism, which is a very powerful 
method for analyzing classical many--body systems with a stochastic dynamic given by a master equation on a
lattice, for a review see \cite{gs}. The Fock--space approach is based on a quantum formalisms for the underlying 
master equation written in terms of creation and annihilation operators. This approach for SFM$\left[ f,d\right] $ will be more transparent by introducing occupation numbers $n_i$ via the relation $\sigma_i = 2 n_i -1 $. Thus, the 
spin down orientation $\sigma_i = -1 $ and up orientation $\sigma _i = 1$ correspond to an empty, $n_i = 0$ and 
a single occupied site with $n_i= 1$, respectively. The SFM$\left[ f,d\right] $ can then be studied in the 
lattice gas representation with excluded volume effects, i.e. the change of a certain configuration 
${\bf n}= \left\{ n_{i}\right\} $ is only possible when the exclusion principle is fulfilled. Following 
Refs. \cite{do,gr,pe,satr,schutr1,schutr2}, the probability distribution $P({\bf n},t)$ can be related to a state 
vector $\left| F(t)\right\rangle $ in a Fock-space according to $P({\bf n},t)=\left\langle {\bf n}|F(t)\right\rangle $ 
and $\left| F(t)\right\rangle =\sum_{{\bf n}}P({\bf n},t)\left| {\bf n}\right\rangle $, respectively, where the basic  vectors $\left| {\bf n} \right\rangle $ can be expressed by second quantized operators. Using this representation, the 
underlying master equation can be transformed into an equivalent evolution equation in a Fock-space, written in the form:  
\begin{equation}
\partial _{t}\left| F(t)\right\rangle =L\left| F(t)\right\rangle\,.  
\label{ev1}
\end{equation}
The dynamical matrix $L({\bf n},{\bf n}^{\prime })$ of the master equation is mapped onto the operator 
$L = L(a,a^{\dagger })$, where $a$ and $a^{\dagger }$ are the annihilation and creation operators, respectively. 
Originally, this transformation had been applied for the Bose case with unrestricted occupation numbers \cite{do,gr,pe}. Here, we consider the case of restricted occupation numbers \cite{satr,schutr1,schutr2}. In order to preserve the restriction of the occupation number in the underlying dynamical equations, the commutation rules of the operators $a$ and $a^{\dagger }$ are chosen as Pauli-operators \cite{satr,gwsp,al} with the commutation rules: 
\begin{equation}
\lbrack a_{i},a_{j}^{\dagger }]=\delta _{ij}(1-2a_{i}^{\dagger }a_{i})\quad
\quad [a_{i},a_{j}]=[a_{i}^{\dagger },a_{i}^{\dagger }]=0\quad \quad
a_{i}^{2}=(a_{i}^{\dagger })^{2}=0\,.
\end{equation}
The master equation of the SFM$\left[ f,d\right] $ can be expressed by the following evolution operator 
\cite{st,schutrFAM} 
\begin{equation}  
L=\sum_{i}\left\{ {G \choose f}^{-1}\sum_{k_{1},...k_{f}}\kappa_{i|k_{1}...k_{f}}N_{k_{1}}N_{k_{2}}...N_{k_{f}}\right\}\, \left[ \beta(a_{i}-N_{i})+\lambda (a_{i}^{\dagger }-(1-N_{i}))\right]\,,  
\label{evolut}
\end{equation}
where $N_{i}=a_{i}^{\dagger }a_{i}$  is the particle number operator which satisfies 
$N_{i}|{\bf n}\rangle = n_{i}|{\bf n}\rangle $. Whereas the temperature dependent flip-rate 
$\lambda $ stands for the process $\downarrow \rightarrow \uparrow $, the rate 
$\beta > \lambda $ moderates the inverse process. The temperature $T$ is introduced in the model by 
$T\sim [\ln (\beta /\lambda )]^{-1}$. The terms in the squared bracket in Eq.~(\ref{evolut}) represent 
a single spin--flip process at lattice site $i$. The product of the particle number operators 
$N_{k_{1}}N_{k_{2}}...N_{k_{f}}$ takes into account the constraints for the flip-process at site $i$. 
The lattice function $\kappa _{i|k_{1} \dots k_{f}}$ is only different from zero if all $f$ lattice sites 
$k_1 \dots k_f$ are different from each other and moreover, all sites are inside the kinetic active zone 
around the lattice point $i$, i.e. all sites $k_{\alpha }$ with $\alpha = 1 \dots f$ satisfy the condition 
$\left| k_{\alpha }-i \right| < R$. In that manner flip processes are only allowed if all the lattice sites 
$k_{\alpha},\,\alpha =1 \dots f $ in the surroundings of site $i$ are occupied. Notice that the restriction number 
$f$ with $f \leq G$, where $G$ is the number of considered neighborhood sites in the environment of the lattice point 
$i$. Introducing the operator  
\[
S=\sum_{k\in E_{i}(R)}N_{k} 
\]
the constraints can be rewritten using some combinatorial manipulations. Then the evolution operator 
Eq.~(\ref{evolut}) reads 
\begin{equation}
L= \sum_{i}{{G \choose f}}^{-1}
{S \choose f} \left[\beta (a_{i}-N_{i})+ {\lambda }(a_{i}^{\dagger }-(1-N_{i}))\right]\,. 
\label{evolut1}
\end{equation}
Obviously, the rate for the flip-process $\sigma =+1\rightarrow \sigma =-1$ becomes $\beta $ for the case 
that all spins are in the up state, independently from the choice of $E_{i}(R)$ and therefore of $G$. 
To proceed further we follow Doi \cite{do} and calculate the average of an arbitrary physical quantity 
$B({\bf n}) $ by using the average of the corresponding operator 
$B(t)=\sum_{{\bf n}}\left| {\bf n}\right\rangle B({\bf n})\left\langle {\bf n}\right| $ 
via\cite{schutr0} 
\begin{equation}
\left\langle B(t)\right\rangle = \sum_{{\bf n}} P({\bf n},t)B({\bf n})= 
\left\langle s\left| B \right| F(t)\right\rangle.  
\label{ev2}
\end{equation}
Here we have used the reference state 
$\left\langle s\right| =\sum_{{\bf n}}\left\langle {\bf n}\right| $. The normalization condition for the 
probability density is included in the condition $\left\langle s|F(t)\right\rangle =1$ with the consequence 
\cite{schutr0}, that the evolution operator fulfills always the relation $\langle s|\,L = 0$.
Note, the special feature of this Fock--space formulation is due to the fact
that the average value is linear in the corresponding state vector in
contrast to quantum mechanics where it is bilinear. In the same way,
correlation functions can be expressed as 
\begin{equation}
\left\langle A(t)B(t^{\prime })\right\rangle =\sum_{{\bf n},{\bf n}^{\prime}}
A({\bf n})P({\bf n},t\,;{\bf n}^{\prime },t^{\prime })B({\bf n}^{\prime
})=\left\langle s\left| A\exp \left\{ L\left( t-t^{\prime }\right) \right\}
B\right| F(t^{\prime })\right\rangle\,.
\label{evolut2}
\end{equation}
Furthermore, because of Eq.~(\ref{ev1}) and Eq.~(\ref{ev2}) the evolution equation
for an arbitrary operator $B(t)$, for example the particle number operator,
is given by\cite{schutr0} 
\begin{equation}
\partial _{t}\left\langle B\right\rangle =\left\langle s\left|
[B,L]\,\right| F(t)\right\rangle\,,  
\label{kine}
\end{equation}
which can be extended immediately in order to get the kinetic equations for time-dependent correlation 
functions. As a general result of the quantum formulation of stochastic processes, all dynamical equations, 
describing the classical problem, are completely determined by the commutation rules of the underlying
operators and the structure of the evolution operator $L$. The Fock-space approach for the master equation allows the analysis of a broad variety of evolution processes, e.g. aggregation, chemical reactions \cite{schutr1,schutr2}, 
non--linear diffusion \cite{schutr3} as well as the spin facilitated kinetic Ising models, compare also \cite{gs}. 
One of the advantages of the Fock--space approach is given by the simple construction principles for each evolution 
operator $L$ on the basis of creation and annihilation operators \cite{schupig}.\\ 
\noindent The evolution equation for the averaged particle number operator, using Eq.~(\ref{kine}) reads 
\begin{equation}
\frac{\partial }{\partial t}\left\langle N_{i}\right\rangle =
{{G \choose f}}^{-1}\left[ \lambda \left( \left\langle 
{{S \choose f}}\right\rangle -\left\langle 
{{S \choose f}}N_{i}\right\rangle \right) -\beta \left\langle 
{{S \choose f}}N_{i}\right\rangle \right]\,.  
\label{g1}
\end{equation}
The evolution equation for the autocorrelation function becomes 
\begin{eqnarray}
\frac{\partial }{\partial t}\left\langle N_{i}(t)N_{i}(t^{\prime
})\right\rangle &=&{{G \choose f}}^{-1}\lambda \left( \left\langle 
{{S(t) \choose f}}N_{i}(t^{\prime })\right\rangle -\left\langle 
{{S(t) \choose f}}N_{i}(t)N_{i}(t^{\prime })\right\rangle \right)  \nonumber \\
&-&{{G \choose f}}^{-1}\beta \left\langle 
{{S(t) \choose f}}N_{i}(t)N_{i}(t^{\prime })\right\rangle\,,  
\label{g2}
\end{eqnarray}
which is valid for $t > t^{\prime }$. Due to the constraints both evolution equations contain higher 
moments of spins at neighboring lattice sites. The evolution equation of these moments contains the next higher correlations, i.e. one obtains an infinitely large hierarchy of evolution equations. In Eq.~(\ref{g1}) the higher 
moments decouple in the equilibrium state, because the Hamiltonian of the SFM$\left[ f,d\right] $
is that of a simple paramagnetic gas. Furthermore, the equilibrium is given by 
$\partial \left\langle N_{i}\right\rangle /\partial t=0$ leading immediately to  
$$
N_{{\rm eq}}= \frac{\lambda}{\lambda +\beta }\,.
$$ 
This simple decoupling fails outside the equilibrium due to the strong kinetic coupling between neighbored 
spins. But in case of diverging size $R$ of the active zone and consequently for $G\rightarrow \infty $ the 
intrinsic self averaging effect due to the infinite number of neighbored spins results again in a decoupling of 
higher correlations.

\subsection{Long range kinetic interaction}

\noindent When the system is subjected to homogeneous initial conditions then the averaged particle number operator 
is independent on the lattice site. Denoting that value by $\langle N_i \rangle = \overline{N}$ 
we obtain for large $G$ ($G \to \infty$) from Eq.~(\ref{g1}) 
\begin{equation}
\frac{\partial \overline{N}}{\partial t}={{G \choose f}}^{-1}{{G\overline{N} \choose f}}
\left( \lambda \left[ 1-\overline{N}\right] -\beta \overline{N}\right)
= \left( \lambda +\beta \right) 
{{G \choose f}}^{-1}{{G\overline{N} \choose f}}\left( N_{{\rm eq}}-\overline{N}\right)\,.  
\label{f1}
\end{equation}
Using the same arguments the evolution equation for autocorrelation function, Eq.~(\ref{g2}), reads 
\begin{equation}
\frac{\partial }{\partial t}\left\langle N(t)N(t^{\prime })\right\rangle ={{G \choose f}}^{-1}
{{G\overline{N}\left( t\right)  \choose f}}\left( \lambda \left[ \overline{N}(t^{\prime })-\left\langle
N(t)N(t^{\prime })\right\rangle \right] -\beta \left\langle N(t)N(t^{\prime
})\right\rangle \right)\,.  
\label{ff}
\end{equation}
The last equation can be simplified by introducing the two-time correlation function 
$H(t,t^{\prime })=\overline{N(t)N(t^{\prime })} -\overline{N}(t)\overline{N}(t^{\prime })$.  
It follows
\begin{equation}
\frac{\partial }{\partial t}H(t,t^{\prime })=-\left( \lambda +\beta \right) {{G \choose f}}^{-1}
{{G\overline{N}\left( t\right)  \choose f}}H(t,t^{\prime })\,.  
\label{f2}
\end{equation}
This equation can be solved explicitly. We get for $t > t^{\prime }$ 
\[
\ln \frac{H(t,t^{\prime })}{H(t^{\prime },t^{\prime })}=-\left( \lambda
+\beta \right) {{G \choose f}}^{-1}\int\limits_{t^{\prime }}^{t}
{{G\overline{N}\left( \tau \right)  \choose f}}d\tau \,.
\]
Using Eq.~(\ref{f1}), we find 
\[
\ln \frac{H(t,t^{\prime })}{H(t^{\prime },t^{\prime })}=-\left( \lambda
+\beta \right) \int\limits_{\overline{N}\left( t^{\prime }\right) }^{%
\overline{N}\left( t\right) }\frac{d\overline{N}}{\lambda -\left( \lambda
+\beta \right) \overline{N}}=\ln \frac{\lambda -\left( \lambda +\beta
\right) \overline{N}(t)}{\lambda -\left( \lambda +\beta \right) \overline{N}%
(t^{\prime })}\,, 
\]
which yields 
\begin{equation}
\frac{H(t,t^{\prime })}{H(t^{\prime },t^{\prime })}=\frac{\overline{N}(t)- 
N_{{\rm eq}}}{\overline{N}(t^{\prime })-N_{{\rm eq}}}\,.  
\label{cw0}
\end{equation}
The correlation function $H(t,t^{\prime })$ depends on $t$ and $t^{\prime}$ with $t^{\prime} < t $. 
Whereas $t$ is the observation time, $t^{\prime }$ can be interpreted as waiting time $t_{W}$. The system 
relaxes from the initial state at $t=0$ into the instantaneous state at $t$ via an intermediate time 
$t^{\prime} = t_W $. It seems to be reasonable to introduce the reduced correlation function $C(\tau, t_W)$ 
by the definition
\begin{equation}
C(\tau ,t_{W})= \frac{H(t_{W}+\tau ,t_{W})}{H(t_{W},t_{W}}\,.
\label{red}
\end{equation}
The reduced correlation function describes the relaxation behavior of the 
SFM$\left[ f,d\right] $.\\ 
\noindent The time dependent decay $\overline{N}(t)$ follows from the solution of Eq.~(\ref{f1}). We start from the 
initial condition $\sigma _{i}= + 1$ for all lattice sites $i$, i.e. we have the initial condition 
$\overline{N}(0)= 1 > N_{{\rm eq}} $. Furthermore, one concludes from Eq.~(\ref{f1}) the validity of 
$1\ge \overline{N}(t)>N_{{\rm eq}}$ for all times. The SFM$\left[ f,d\right] $
approaches the thermodynamic equilibrium state, if $GN_{{\rm eq}} > f$,
otherwise the system reaches the stationary state $N_{{\rm st}}=f/G>N_{{\rm eq}}$.

\section{Results}

\subsection{Relaxation into the equilibrium state}

\begin{figure}[!ht]
\centering
\includegraphics[scale=0.8]{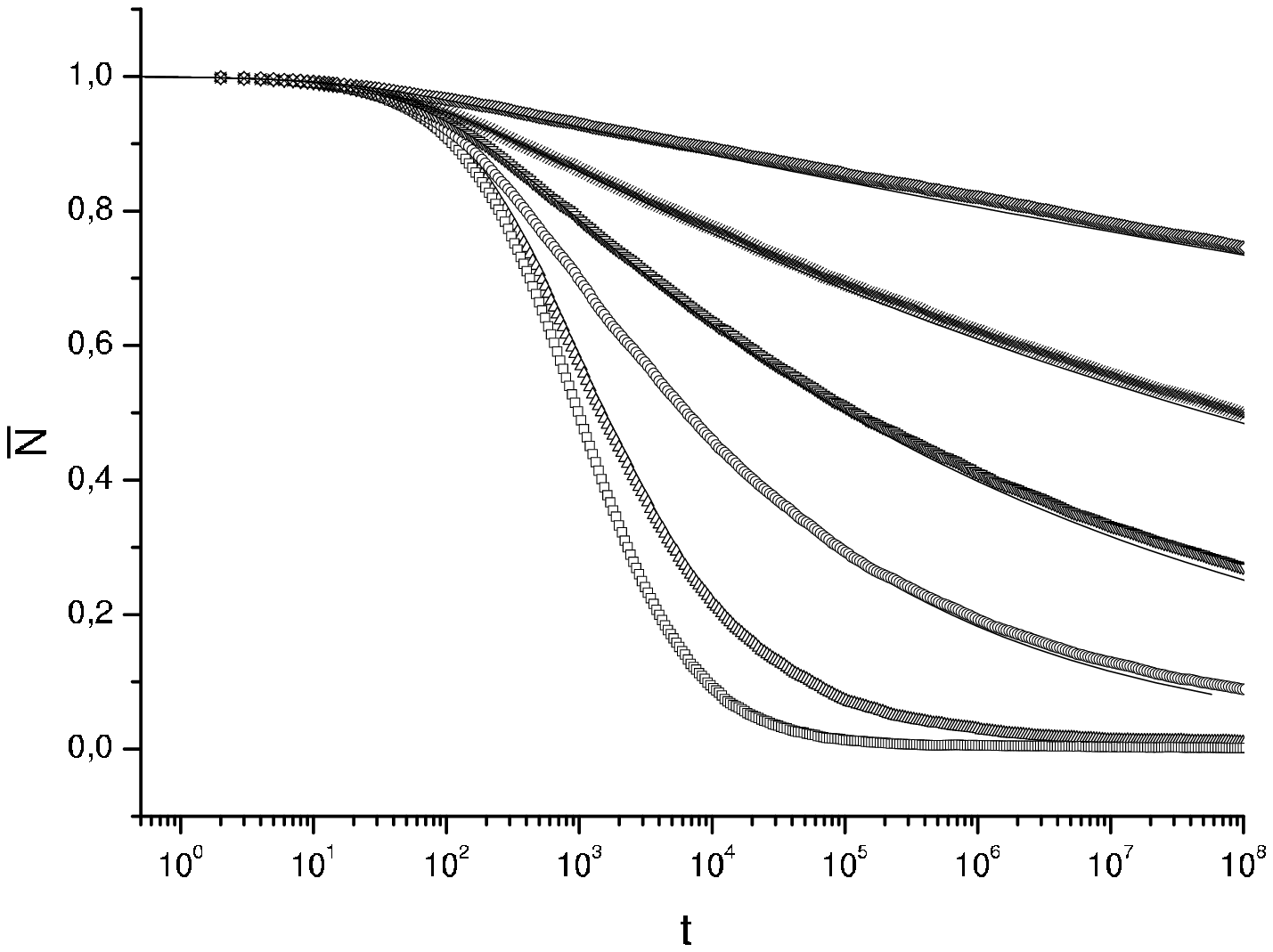}
\caption{$\overline{N}(t)$ for various $f$ and fixed $G=200$ in comparison with
the analytical mean field results (straight lines). The temperature is given
by $T=(3\ln 10)^{-1}$. The values of $f$ are $f=1$ (squares), $f=2$ (up
triangles), $f=5$ (circles), $f=10$ (down triangles), $f=20$ (crosses) and $%
f=50$ (diamonds) }
\label{Fig.1}
\end{figure}

\begin{figure}[!ht]
\centering
\includegraphics[scale=0.8]{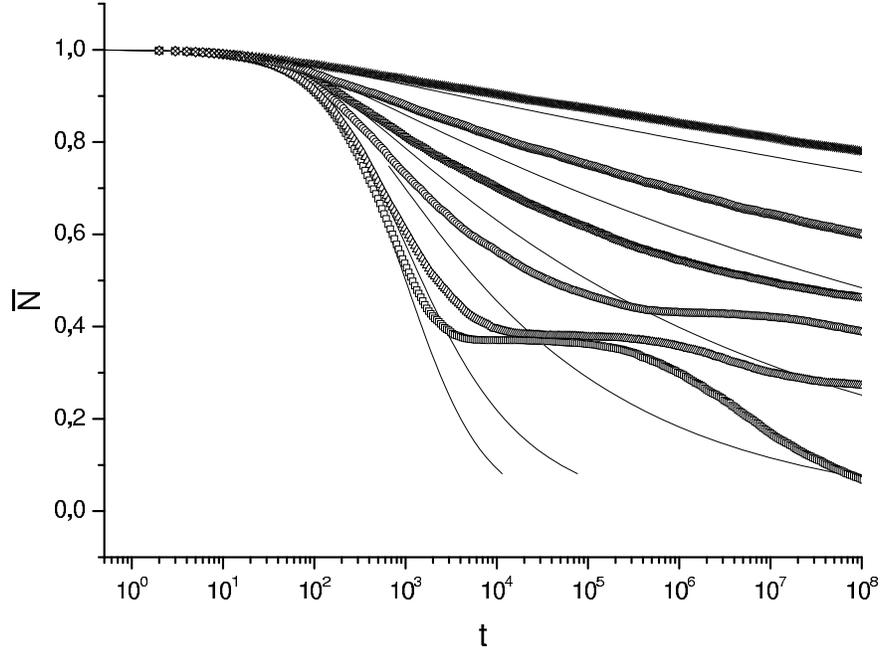}
\caption{$\overline{N} (t)$ for various $f$ and fixed ratio $f/G=1/2$ in
comparison with the analytical mean field results (straight lines). The
temperature is given by $T=(3 \ln 10)^{-1}$. The values of $f$ are $f=1$
(squares), $f=2$ (up triangles), $f=5$ (circles), $f=10$ (down triangles), $%
f=20$ (diamonds) and $f=50$ (crosses) }
\label{Fig.2}
\end{figure}

\noindent Let us firstly analyze the behavior for finite $f$. In this case we obtain for 
$f/G\rightarrow 0$, i.e. $G\rightarrow \infty $ and consequently 
\begin{equation}
{{G \choose f}}^{-1}{{G\overline{N} \choose f}}\rightarrow \overline{N}^{f}\,.  
\label{curr}
\end{equation}
Inserting this expression in Eq.~(\ref{f1}) the evolution equation reads now (compare also \cite{st}) 
\begin{equation}
\frac{\partial \overline{N}/N_{{\rm eq}}}{\partial t}=\left( \lambda +\beta
\right) N_{{\rm eq}}^{f}\left( \frac{\overline{N}}{N_{{\rm eq}}}\right)
^{f}\left( 1-\frac{\overline{N}}{N_{{\rm eq}}}\right)\,.  
\label{cvt}
\end{equation}
From here, we obtain the implicit solution 
\begin{equation}
-\ln \left( \frac{\overline{N}-N_{{\rm eq}}}{1-N_{{\rm eq}}}\right) +\ln 
\overline{N}+\sum_{k=2}^{f}\frac{N_{{\rm eq}}^{k-1}}{k-1}\left[ 1-\frac{1}
{\overline{N}^{k-1}}\right] =N_{{\rm eq}}^{f}\left( \lambda +\beta \right) t\,.
\label{sol1}
\end{equation}
The sum can be rewritten by using the Lerch-$\Phi $-function 
\begin{equation}
\Phi (z,a,b)=\sum_{n=0}^{\infty }\frac{z^{n}}{(b+n)^{a}}\,.
\end{equation}
in the form as follows
\begin{equation}
\frac{1}{\overline{N}^{f}}\Phi \left( \frac{N_{{\rm eq}}}{\overline{N}},1,f\right) =\Phi (N_{{\rm %
eq}},1,f)+\left( \lambda +\beta \right) t\,.  
\label{CCC}
\end{equation}
The function $\overline{N}(t)$ decreases monotonously from $\overline{N}(0)=1 $ to 
$\overline{N}(\infty )=N_{{\rm eq}}$. The short time behavior can be calculated by a series expansion of 
Eq.~(\ref{sol1}). Up to second order in $t$ we get for $t \to 0$ 
$$
\overline{N}=1-\beta t+(f\beta +\lambda +\beta )\beta\,t^{2}/2+o(t^{3})\,. 
$$ 
This short time regime is valid for $t < (\beta f)^{-1}$. The early stage, described by the first order of the
expansion, depends only on the kinetic constant quantifying downward spin flips. Backward flips and kinetic 
restrictions becomes relevant at later stages, because the corresponding parameter $f$ and the kinetic rate appear
firstly in the second order term.\\ 
\noindent The final time regime is given by $\overline{N}\approx N_{{\rm eq}}$. Obviously, the first logarithmic term in 
Eq.~(\ref{sol1}) is the dominate one. Then we find the asymptotic solution 
\begin{equation}
\overline{N}(t)-N_{{\rm eq}}= (1 - N_{{\rm eq}}) \exp \left\{ -\left( \lambda +\beta \right) N_{{\rm eq}}^{f}t\right\}\,.  
\label{expo}
\end{equation}
From Eq.~(\ref{sol1}) follows, that this exponential decay regime becomes relevant for 
$t\gg f^{-1}N_{{\rm eq}}^{-f}(\lambda +\beta )^{-1}$.\\ 
\noindent Further, the evolution equation allows a pronounced intermediate regime resulting for 
$N_{{\rm eq}}\ll 1$ and large $f$. In this case we find for $1\gg \overline{N}(t)\gg N_{{\rm eq}}$ a power 
law behavior in form of
\begin{equation}
\overline{N}\approx \left[ 1+f\beta t\right] ^{-1/f}\sim t^{-1/f}\,.
\label{int}
\end{equation}
\noindent For infinite large $f$, but $f/G\rightarrow 0$ both, the description of the decay of the initial 
regime and the intermediate regime may be combined to the general relation  
\begin{equation}
\overline{N}=1-\frac{1}{f}\ln \left[ 1+f\beta t\right]   
\label{log}
\end{equation}
for all finite times $t\ll {\rm e}^{f}$. This result can be obtained directly from 
Eq.~(\ref{cvt}) considering $\overline{N}(t)\approx 1$.\\ 
\noindent In case of a finite ratio $f/G$ and $G\rightarrow \infty $ we get for $\overline{N}\gg
f/G$ instead of (\ref{curr}) the relation 
\begin{equation}
{{{G \choose f}}}^{-1}{{{G\overline{N} \choose f}
}}\rightarrow \left( \overline{N}\exp \left\{ \frac{f}{2G}(1-\overline{N}\,,
^{-1})\right\} \right) ^{f}  
\label{murr}
\end{equation}
which indicates a further slowing-down of the relaxation with decreasing
ratio $f/G$ and increasing $\overline{N}$. An explicit expression for the time evolution can be derived 
by inserting Eq.~(\ref{murr}) in Eq.~({f1}) and making the ansatz $\overline{N}(t) = 1 - y(t)$ with 
$y \ll \beta /(\lambda + \beta ) < 1 $. It results
\begin{equation}
\overline{N}(t) \simeq 1 - \frac{2G}{f^2}\ln\left[\frac{\beta f^2}{2G}t\right]\,.
\label{int2}
\end{equation}
\noindent While the case of $R\rightarrow \infty $ can be solved analytically, the
analysis for finite $R$ requires numerical methods. Here, we use a standard Monte-Carlo Simulation 
\cite{schu1,schu3} realized on a $d=1$ dimensional lattice. Fig.~\ref{Fig.1} shows the decay $\overline{N}(t)$ for 
fixed $G$ and various $f\ll G$ in comparison with the predictions obtained from the mean field theory. As
expected, the accuracy of the mean-field solution Eq.~(\ref{CCC}) increases with decreasing ratio $f/G$. This 
phenomenon is partially justified by the relation Eq.~(\ref{murr}) which slows down the decay described by 
Eq.~(\ref{CCC}). Additionally, an increasing ratio $f/G$ requires an increasing number of
spin flips to collect $f$ up-spins in the environment of the lattice site $i$
before the spin $\sigma _{i}$ is able to flip. While the first effect is contained in the general mean field 
approach given by Eq.~(\ref{f1}), the cooperative phenomenon is not taken into account by the mean field
concept. Fig.~\ref{Fig.2} shows the decay of $\overline{N}(t)$ for various $f$ and $G=2f$.
Because the temperature $T\sim (\ln \beta /\lambda )^{-1}$ is very small, the initial regime is partially 
comparable with the situation for $T=0$. Here the dynamics stop at a non ergodic state. The approach into the
non ergodic state can be demonstrated by rigorous results \cite{schutrFAM},
in this connection see also \cite{palmer}. In particular, for $f=1$ and $d=1$
the decay of the ordered phase $\overline{N}=1$ is given by 
\begin{equation}
\overline{N}\left( t\right) =\exp \left\{ \exp \left( -\beta t\right)-1\right\}\,.  
\label{phim}
\end{equation}
In the long time limit this results in a pronounced slowing-down leading to
a non ergodic behavior which is manifested in that for $t\rightarrow \infty $
the quantity $\overline{N}(t)$ remains finite with $\overline{N}(\infty)=e^{-1}$.\\ 
\noindent For all finite temperatures the non ergodic state is unstable and decays by
cooperative rearrangements of the spin configurations. For sufficiently long
times the system approaches the equilibrium state, $\overline{N}(t\rightarrow \infty )=N_{{\rm eq}}$. This 
slow process is well described by a mapping of the kinetically constrained spin flips onto an auto catalytic
reaction diffusion process on a one-dimensional lattice \cite{schutrFAM},
which offers a slightly stretched exponential behavior.\\ 
\noindent The decay $\overline{N}(t)$ approaches again the mean field result, obtained in  
Eq.~(\ref{CCC}) for sufficiently large $f$, which indicates the decreasing relevance of
cooperative processes for large $f$. However, a larger difference between
the solution Eq.~(\ref{CCC}) and the numerical result remains. This difference
occurs for all $\overline{N}(t)<1$ and it is related to the fact that Eq.~(\ref{curr}) is valid only for 
$f/G\rightarrow 0$. For finite $f/G$, Eq.~(\ref{curr}) is an approximation which becomes exact only for 
$\overline{N}(t)\rightarrow 1$, see Eq.~(\ref{murr}). Therefore, the decay limit Eq.~(\ref{log})
obtained for $f\rightarrow \infty $ is also valid for all finite values $f/G$ with $t\ll {\rm e}^{f}$.

\subsection{Correlation function}

\begin{figure}[!ht]
\centering
\includegraphics[scale=0.8]{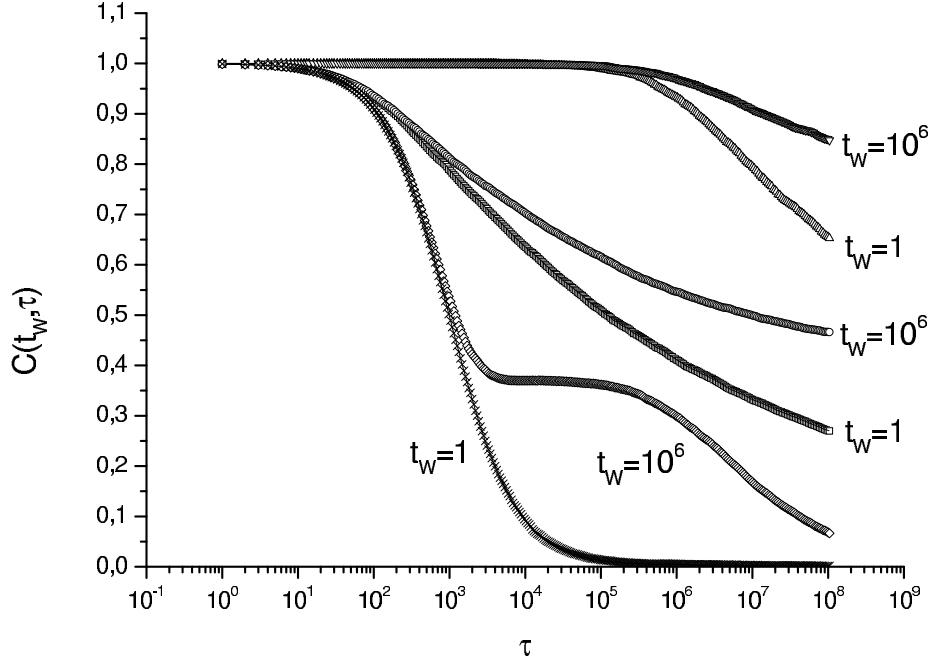}
\caption{ Correlation function $C( t_W , \protect\tau )$ computed from 
Eq.~(\ref{cw0}) using the numerically determined function $\overline{N}$ (straight
lines) and obtained from direct numerical simulations for $t_W=1$ and $%
t_W=10^6$ ($f=10$, $G=200$: squares and circles), ($f=10$, $G=20$: up and down triangles), 
($f=1$, $G=2$: crosses and diamonds) }
\label{Fig.3}
\end{figure}

\begin{figure}[!ht]
\centering
\includegraphics[scale=0.8]{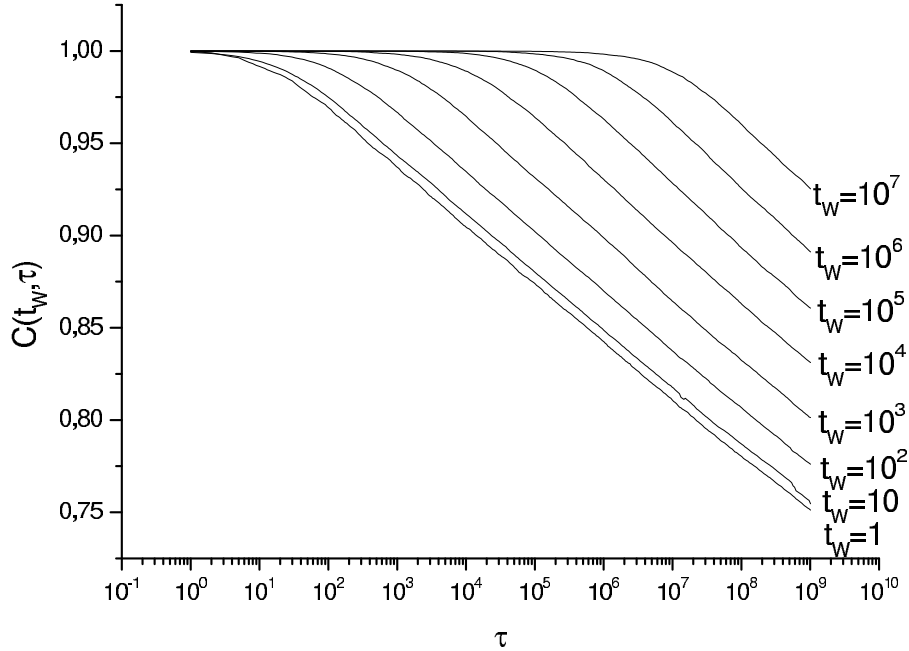}
\caption{ Correlation function $C( t_W , \protect\tau )$ for fixed $f=50$
and $G=100$ and different waiting times $t_W =10^n$ with $n=0 \dots 6$. }
\label{Fig.4}
\end{figure}

\begin{figure}[!ht]
\centering
\includegraphics[scale=0.8]{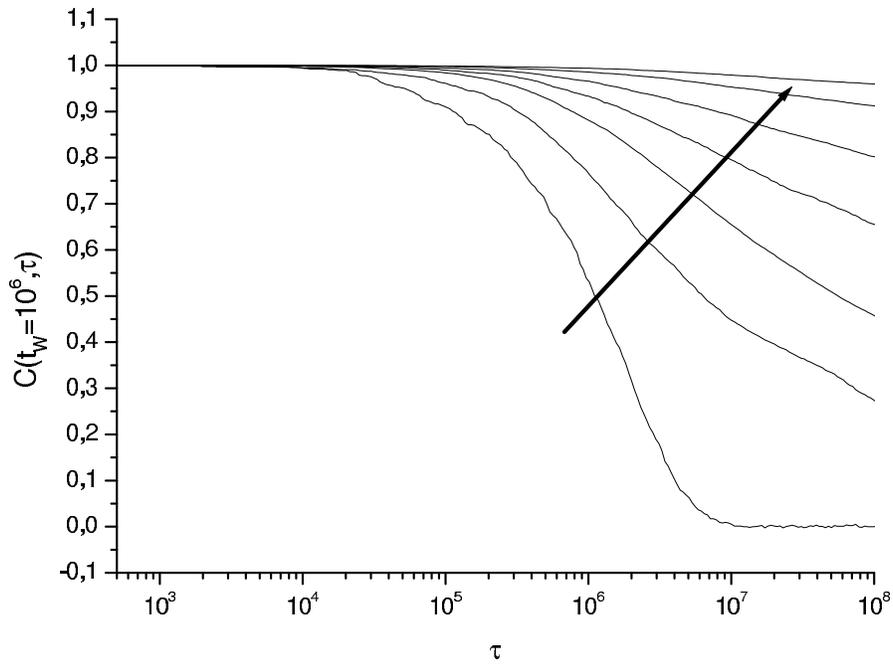}
\caption{ Correlation function $C( t_W , \protect\tau )$ for fixed $G=200$
and waiting time $t_W=10^6$ and $f=1,\,2,\,5,\,10,\,20,\,50$ and $100$. The
value of $f$ increases in the direction of the arrow. }
\label{Fig.5}
\end{figure}

\noindent The knowledge of $\overline{N}(t)$ allows us to compute the reduced
correlation function $C(\tau ,t_{W})$, defined by Eq.~(\ref{red}). The decay of this correlation
function is also characterized by different regimes.\\
(i) For a short waiting time $t_{W}\rightarrow 0$ and short time scale $\tau $ we find an universal relation 
for the reduced correlation function, namely
\begin{equation}
C(\tau ,t_{W})=1-\left( \lambda +\beta \right) (1-\beta ft_{W})\tau\,.
\end{equation}
Note, that a general analysis of Eq.~(\ref{g1}) and Eq.~(\ref{g2})
shows that this quantity is independent of the interaction range $R$.\\
(ii) The limit of a long waiting time $t_{W}\rightarrow \infty $ leads always to 
\begin{equation}
C(\tau ,t_{W})=\exp \left\{ -\left( \lambda +\beta \right) N_{{\rm eq}}^{f}\tau \right\}\,.
\end{equation}
The independence of $t_{W}$ indicates the proximity to the thermodynamic equilibrium.\\
(iii) If the time scale increases, i.e. $\tau \rightarrow \infty $, we obtain
\begin{equation}
C(\tau ,t_{W})=\exp \left\{ -\left( \lambda +\beta \right) N_{{\rm eq}}^{f}(\tau +t_{W}^{*})\right\}
\end{equation}
with 
\[
t_{W}^{*}=t_{W}-\frac{\ln (\overline{N}(t_{W})-N_{{\rm eq}})}{\left( \lambda
+\beta \right) N_{{\rm eq}}^{f}}\,. 
\]
(iv) Within a short or intermediate time scale $\tau $ and an intermediate waiting time $t_{w}$ 
the reduced correlation function offers a characteristic aging behavior manifested by 
\begin{equation}
C(\tau ,t_{W})\sim \left[ 1+\frac{\tau }{t_{W}}\right] ^{-1/f}\,.
\end{equation}
\noindent The crossover to infinite $f$ shows, that the decay processes freeze in. This
means, the short time regime is restricted in terms of $t_{W},\,\tau<(\beta f)^{-1}$. In the intermediate time 
regime the correlation function becomes 
\begin{equation}
C(\tau ,t_{W})\sim 1-\frac{1}{f(1-N_{{\rm eq}})}\ln \left( 1+\frac{\tau }{t_W}\right)\,.  
\label{log2}
\end{equation}
The result is valid up to $\tau /t_{W}\sim {\rm e}^{f}$. The correlation function $C(\tau , t_W)$, characterizing 
the aging behavior, reveals a logarithmic decay with an extreme small prefactor. Thus, the intermediate 
time regime dominates the behavior of the system for all finite times $\tau $ and $t_{W}$.\\ 
\noindent Our numerical investigations support these mean field results. Fig.~\ref{Fig.3} shows
various correlation functions directly obtained from the numerical
simulations in comparison with the correlation functions Eq.~(\ref{red}) computed from the
numerically determined decays $\overline{N}(t)$ using Eq.~(\ref{cw0}). As
discussed above in the context of the numerical analysis of the decay
function, significant differences appear only for small $f$.\\ 
\noindent Fig.~\ref{Fig.4} shows the dependence of the correlation function on the waiting time 
$t_{W}$ for $f/G=1/2$ and $f=50$. We obtain a typical crossover from an
exponential decay, independent on the waiting time to a power law like behavior which scales with the waiting 
times $t_{W}$. This behavior agrees again with the mean field approach.\\
\noindent Finally, we demonstrate the dependence of the correlation function on the
restriction number $f$ for a fixed waiting time, see Fig.~\ref{Fig.5}. As discussed
above, we observe a continuous crossover to a logarithmic decay for
sufficiently large $f$, suggested by Eq.~(\ref{log2}).

\section{Conclusions}

\noindent We have studied the influence of the restriction number $f$ and the range of
the nearest environment on the kinetics of the SFM$\left[ f,d\right] $. To
this aim the original model was extended by an additional parameter $G$
describing the number of kinetically neighbored spins around a given lattice
site $i$. All spins of this environment have the same influence on the flip
probability of the spin $\sigma _{i}$. We obtain a pronounced slowing down
of the relaxation behavior and the correlation functions for an increasing
ratio $f/G$ and fixed $f$ as well as for increasing $f$ and fixed ratio $f/G$. For sufficiently 
large $f$, a logarithmic decay of the spin-spin
correlation function was observed. Furthermore, in the case of an infinite
kinetic interaction range $G$, the correlation functions in the kinetic
equations decouple and the mean field approach becomes an exact theory. This
approach allows the identification of various regimes. While, the initial
regime is described by universal relations which are valid also if the
decoupling procedure fails, the final regime is always an exponential decay
describing the relaxation into the equilibrium state. The intermediate
regime can be approached by power laws for the correlation functions, which
are significantly controlled by the waiting times.\\ 
\noindent For finite $G$ we observe a further slowing down due to the fact, that
cooperative processes become relevant. These cooperative rearrangements are
necessary to collect $f$ up-spins in the environment of the lattice site $i$
so that the spin $\sigma _{i}$ is able to flip. This effect appears
especially for small kinetic ranges $G$ and large ratios $f/G$. In the case
of increasing $G$ the kinetically active environment $E_{i}$ of a given
lattice site $i$ becomes more and more representative for the actual spin
configuration of the whole system and the mean field solutions converge to
exact results. 

\begin{acknowledgments}
This work has been supported by the DFG: SFB 418 and SFB 569
\end{acknowledgments}

\newpage

\end{document}